\documentclass[12pt,preprint]{aastex}

\newcommand{\bV}{\mathbf{V}}
\newcommand{\bB}{\mathbf{B}}

\newcommand{\bJ}{\mathbf{J}}
\newcommand{\rot}{\mathbf{\nabla} \times}

\shorttitle{Solar wind small scale inertial range}
\shortauthors{Alexandrova et al.}

\begin{document}
\title{Small scale energy cascade of the solar wind turbulence}
\author{O. Alexandrova}
\affil{LESIA/CNRS, Observatoire de Paris, 5 pl. J. Janssen, 92195
Meudon, France}
\author{V. Carbone, P. Veltri}
\affil{Dip.~Fisica/CNISM, Ponte P. Bucci, Cubo 31C, 87036 Rende
(CS), Italy}


\and
\author{L. Sorriso--Valvo}
\affil{LICRYL/CNR, Ponte P. Bucci, Cubo 33B, 87036 Rende (CS),
Italy}



\begin{abstract}
Magnetic fluctuations in the solar wind are distributed according to
Kolmogorov's power law $f^{-5/3}$ below the ion cyclotron frequency
$f_{ci}$. Above this frequency, the observed steeper power law is
usually interpreted  in two different ways:  a dissipative range of
the solar wind turbulence or another turbulent cascade, the nature
of which is still an open question. Using the Cluster magnetic data
we show that after the spectral break the intermittency increases
toward higher frequencies, indicating the presence of non-linear
interactions inherent to a new inertial range and not to the
dissipative range. At the same time the level of compressible
fluctuations  raises. We show that the energy transfer rate and
intermittency are sensitive to the level of compressibility of the
magnetic fluctuations within the small scale inertial range. We
conjecture that the time needed to establish this inertial range is
shorter than the eddy-turnover time, and is related to dispersive
effects. A simple phenomenological model, based on the compressible
Hall MHD, predicts the magnetic spectrum $\sim k^{-7/3+2\alpha}$,
which depends on the degree of plasma compression $\alpha$.
\end{abstract}
\keywords{solar wind turbulence, Hall MHD} \maketitle

\section{Introduction} \label{Intro}

Solar wind, which is highly turbulent, represents a unique
opportunity  to investigate turbulence in natural plasmas via
\emph{in situ} measurements \cite{sw2,noi}. In non-magnetized
fluids, where the energy injection scale is far from the dissipation
one, the intermediate scales (inertial range) are described by the
universal power law Kolmogorov spectrum $k^{-s}$ with $s=5/3$. This
law depends neither on the energy injection nor on the energy
dissipation processes.

In the solar wind, and in the interplanetary space in general, the
mean free path roughly corresponds to the Sun-Earth distance and the
usual dissipation via collisions is negligible. At the same time, in
a  magnetized plasma there is a number of characteristic space and
temporal scales. Investigating solar wind turbulence at these scales
is then a challenging topic from the point of view of basic plasma
physics. Here we will focus our discussion on the ion scales, namely
the ion inertial length $\lambda_i = c/\omega_{pi}$ and the ion
cyclotron frequency $f_{ci}= eB/m_i$. At these scales the fluid-like
approximation of plasma dynamics, the usual Magnetohydrodynamic
(MHD) description, breaks down in favor of a more complex
description of plasma.

Solar wind turbulent spectrum of magnetic field fluctuations follows
a  $\sim f^{-5/3}$ power law below the ion cyclotron frequency
$f_{ci}$. For $f>f_{ci}$ the spectrum steepens significantly, but is
still described by a power law $f^{-s}$, with $s \in (2-4)$
\citep{Leamon98,smith06}.

\emph{In situ} solar wind measurements provide time series data,
i.e. information on frequency in Fourier space. How to get any
information on wave vectors? For fluctuations with velocities much
smaller than the plasma bulk velocity $V$, the Taylor hypothesis is
valid: the observed  variations on a time scale $\delta t$
correspond to variations on the spatial scale $\delta r=V \delta t$.
Therefore there is a direct correspondence between $f$ and $k$
spectra. 
If the solar wind turbulence was a mixture of linear wave modes,
Taylor hypothesis would be well verified only below the spectral
break: the bulk velocity is superalfv\'enic ($V>V_A$, where $V_A$
being the Alfv\'en speed) and so the low frequency fluctuations can
be considered as frozen in plasma. However, whistler waves (with $f
> f_{ci}$ and phase speed $V_{\varphi} > V_A$) do not satisfy the
Taylor hypothesis assumption. In this study we assume that in the
solar wind there are no whistler waves above the spectral break
frequency. This last assumption is supported by results recently
obtained in the Earth's magnetosheath \citep{Mangeney2006}.

Using the Taylor hypothesis the observed solar wind spectrum below
the break is usually attributed to the Kolmogorov's spectrum $\sim
k^{-5/3}$. Above the spectral break, the spectral steepening, $\sim
k^{-3}$, can be interpreted in two different ways. Some authors
associate it to the dissipation range
\citep{Leamon98,Leamon99,Leamon2000,Bale2005,smith06}. Others
suggest that after the spectral break another turbulent cascade
takes place \citep{Biskamp1996,Ghosh96,stawicki,Li01,Galtier06}.

In ordinary fluid flows the dissipation range is described by an
exponential function \citep{frisch}. While in the solar wind, a
well-defined power law is observed after the break point and not an
exponential. Note that power spectra, i.e. the second order
statistics, completely describe Gaussian, or statistically
independent, fluctuations. However, as is well known, fluctuations
cannot be described by a Gaussian statistics in the low-frequency
part of the solar wind turbulence \citep{noi}. Deviations from
Gaussianity, i.e. intermittency \citep{frisch}, may be quantified by
the flatness, the forth-order moment of fluctuations.

In this paper we investigate the nature of magnetic fluctuations in
the high frequency range of the solar wind turbulence, that is
usually called dissipation range. We find that in this range the
flatness increase with frequency. This is similar to what is going
on in the low-frequency range.  The presence of intermittency
together with the well defined power law in the high-frequency part
of the spectrum suggests another turbulent cascade rather than a
dissipation range. This small scale cascade is observed to be much
more compressible than the Kolmogorov-like inertial range, in
agreement with the previous observations by the Wind spacecraft
\citep{Leamon98}. We show that the energy transfer rate and
intermittency are sensitive to the level of compressibility of the
turbulent fluctuations in this range. Finally, we propose a simple
phenomenological model, based on the compressible Hall MHD, which
allows to explain the observed range of the spectral indices in the
high frequency part of the solar wind spectrum.

\section{Turbulent spectra and intermittency}

In the present study we analyze the Cluster magnetic field data up
to $12.5$ Hz (or $100f_{ci}$).  We use 57 minutes of data during the
interval 22:35:00--23:32:00 UT on 5 April 2001, when Cluster was in
the solar wind not connected to the bow-shock. During this period
the interplanetary magnetic filed is $B\simeq 7$ nT, the ion density
is $n_i\simeq 3$ cm$^{-3}$, the plasma bulk velocity is $V\simeq
540$ km/s, the ion temperature is $T_i \simeq 33$~eV, the Alfv\'en
speed is $V_A=B/\sqrt{\mu _0 m_in_i}\simeq 90$ km/s, the ion skin
depth $\lambda_i$ is about $130$~km, the ion Larmor radius is
$\rho_i \simeq 80$~km and the ion plasma beta is $\beta_i=2\mu _0
p_i/B^2\simeq 0.8$, $p_i$ being the pressure of the ions. All the
data are from Cluster spacecraft one, $B$ is determined from the
fluxgate magnetometer (FGM) \citep{Balogh2001}, the other plasma
parameters are provided by CIS/HIA instrument \citep{CIS2001}.
Magnetic field fluctuations are measured by the search coils (SC) of
the STAFF experiment with $0.04$~s time resolution
\citep{Cornilleau2003}. This instrument\footnote{STAFF-SC measurs
magnetic fluctuations above 0.1~Hz. As the Cluster spin frequency is
0.25~Hz, there are artefacts at $0.15$ and $0.35$~Hz in the solar
wind. We note that in the magnetosheath, where the level of magnetic
fluctuations is much higher than in the solar wind, this problem is
almost unnoticeable.} operates in the frequency domain
$(0.35-12.5)$~Hz. We use FGM data to resolve the frequencies below
$0.35$~Hz. Results from Cluster have been compared with data from
the Helios~2 satellite within the range $(0.001-1)f_{ci}$. These
data are selected within a high-speed stream $V\simeq600$~km/s when
the satellite orbited at 0.96 AU \citep{Bruno2003}.


In order to analyze the magnetic field fluctuations in the solar
wind we use the Morlet wavelet transform, defined as
\begin{equation}
\mathcal{W}_i(\tau,t) = \sum_{j=0}^{N-1} B_i(t_j) \psi [(t_j- t)/\tau]
\end{equation}
which represents the transform of the $i$-th component of the
magnetic field $B_{i}(t_j)$, a data time series with equal time
spacing $\delta t$ and $j=0,...,N-1$. Here $\tau$ is a time scale
(the corresponding frequency is $f=1/\tau$), while $\psi (u)=
2^{1/2}\pi ^{-1/4} \cos (\omega_0 u) \exp [-u^2/2]$ is the Morlet
wavelet, where $\omega_0=6$. The square of a wavelet coefficient
gives a ``quantum of energy" of magnetic fluctuations on a scale
$\tau$ at a time $t$. Thus we define the power spectral density of
$B_i$ as
\begin{equation}\label{eq:PSD}
\mathcal{S}_i[\textrm{nT}^2/\textrm{Hz}]=\frac{2\delta
t}{N}\sum_{j=0}^{N-1}|\mathcal{W}_i(\tau,t_j)|^2.
\end{equation}

\begin{figure}
\epsscale{.80} \plotone{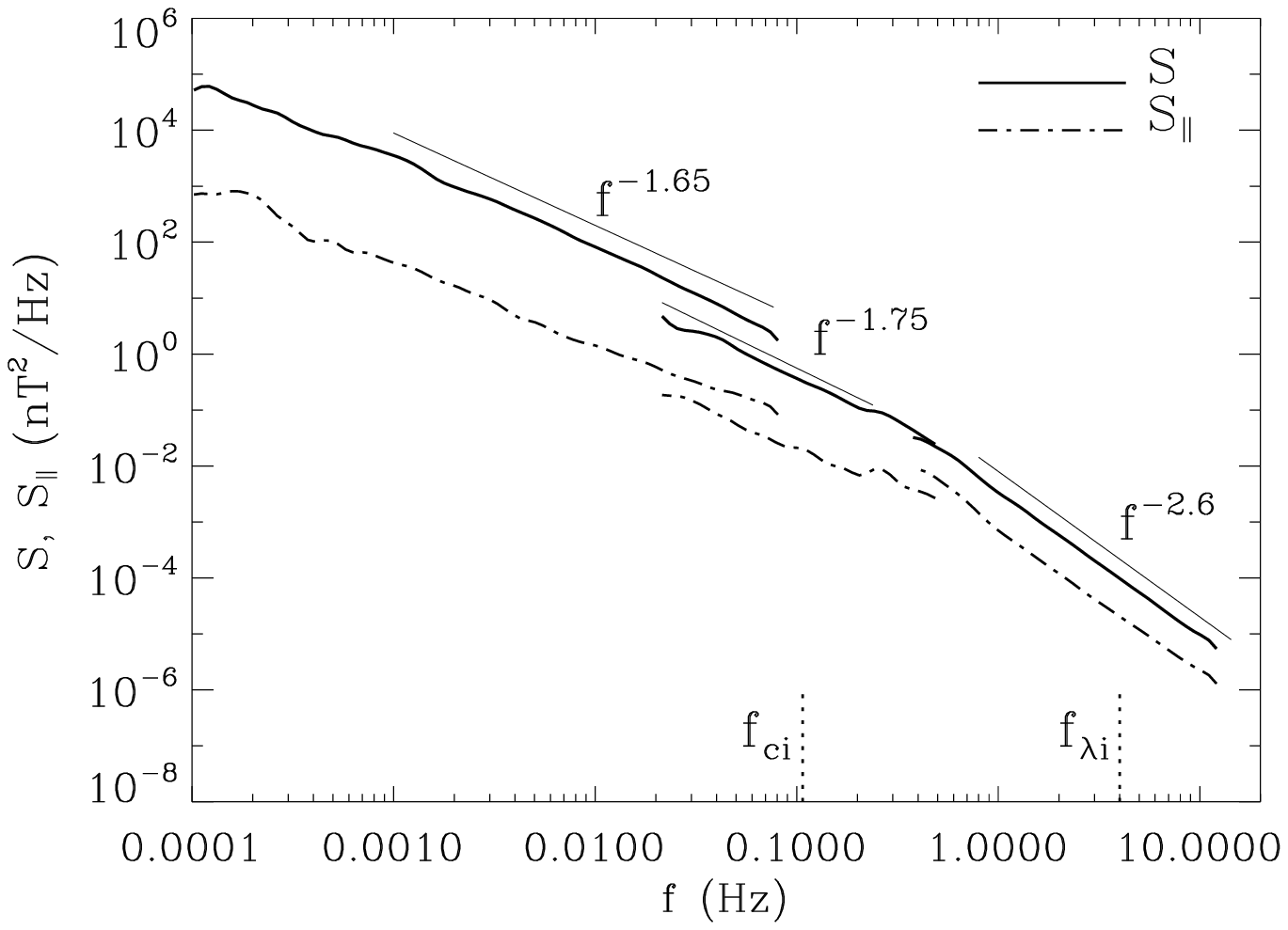} \caption{The total magnetic field
power spectral density $\mathcal{S}$ and the spectrum of
compressible magnetic fluctuations $\mathcal{S}_{\|}$ (dashed-dotted
line) measured by Helios 2 (up to $0.08$ Hz) and by Cluster (up to
$12.5$~Hz). Straight lines refer to power law fits. Vertical dotted
lines indicate the ion cyclotron frequency $f_{ci}$ and the
Doppler-shifted ion inertial length $f_{\lambda_i}$.}
\label{fig:spec}
\end{figure}

Fig.~\ref{fig:spec} shows the total power spectrum density of the
magnetic fluctuations $\mathcal{S}= \sum_{i=x,y,z} \mathcal{S}_i$
(solid lines) and the spectrum  of magnetic field modulus
fluctuations $\delta |{\bf B}|$ (dashed-dotted lines), which
corresponds to the spectrum of parallel fluctuations of magnetic
field $\mathcal{S}_{\|}$. In agreement with previous observations,
the solar wind spectra follow well defined power laws with a break
around $f_b\simeq 0.3$~Hz, in the vicinity of $f_{ci}\simeq 0.1$~Hz.
Taking into account the Taylor hypothesis, $f_b$ corresponds to a
spatial scale $V/f_b \simeq 1800$~km $\simeq 15 \lambda_i$ and to
the normalized angular wavenumber $k \lambda _i \simeq 0.4$ ($k
\rho_i \simeq 0.3$). Below the break, the Helios spectrum
$\mathcal{S} \sim f ^{-1.65}$ is fitted roughly by Kolmogorov's law.
The Cluster FGM spectrum covers a $(0.02-0.5)$~Hz frequency range,
the lower resolved frequency here is determined by the length of the
considered time period, 57 minutes, and by a cone of influence of
the Morlet wavelet transform  \citep{Torrence1998}. Even if the
frequency range is not that large, this spectrum can be described by
a power law $\mathcal{S} \sim f ^{-1.75}$. Above the break, the
spectrum follows a $\mathcal{S} \sim f ^{-2.6}$ law: the spectral
index 2.6 lies in the usual observed range, between 2 and 4
\citep{smith06}. As regards the spectrum $\mathcal{S}_{\|}$, it is
remarkable, that after the spectral break, the compressible magnetic
fluctuations become much more important than in the low frequency
part of the spectrum. This can be quantified by the ratio $\langle
\mathcal{S}_{\|}\rangle/\langle\mathcal{S}\rangle$, that is
$0.02-0.05$ in the law frequency  part and $0.26$ in the high
frequency part of the spectrum. This means that nearly 30\% of the
magnetic fluctuations above the spectral break  are compressible.

\begin{figure}
\epsscale{.80} \plotone{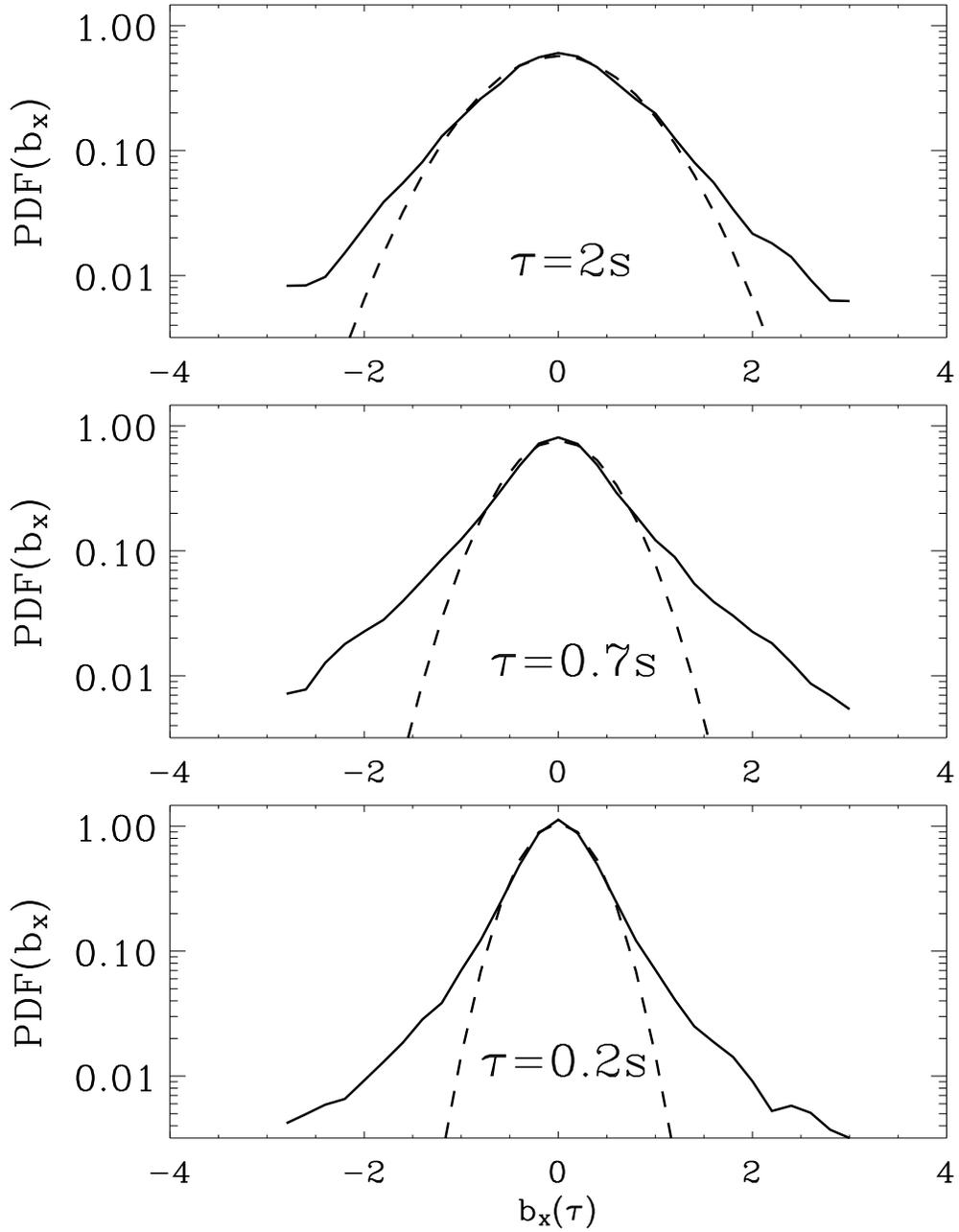}\caption{The probability distribution
functions (PDFs) of the normalized variable $b_x(\tau)$ at three
time scales: $\tau = 2$ s (upper panel), $\tau = 0.7$ s (middle
panel), and  $\tau = 0.2$ s (bottom panel). For each panel a
Gaussian function is plotted for reference as a dashed curve.}
\label{fig:pdfs}
\end{figure}

Let us now consider statistical properties of turbulent fluctuations
in the high-frequency part of the spectrum. Fig.~\ref{fig:pdfs}
shows the  probability distribution functions (PDFs) of the
normalized  variables $b_x(\tau)=\mathcal{W}_x(\tau,t)/\sigma_x$,
with the standard deviation $\sigma_x=\left\langle
\mathcal{W}_x(\tau,t)^2 \right\rangle^{1/2}$, at three different
time scales $\tau$; the dashed lines indicate the corresponding
Gaussian fits. Apparently, the smaller the scale, the more the PDFs
deviate from Gaussian. The PDFs of the other components of the
magnetic field present a similar behaviour. The dependence of
normalized PDFs on  scale is a signature of intermittency
\cite{frisch,noi}.

\begin{figure}
\epsscale{.80} \plotone{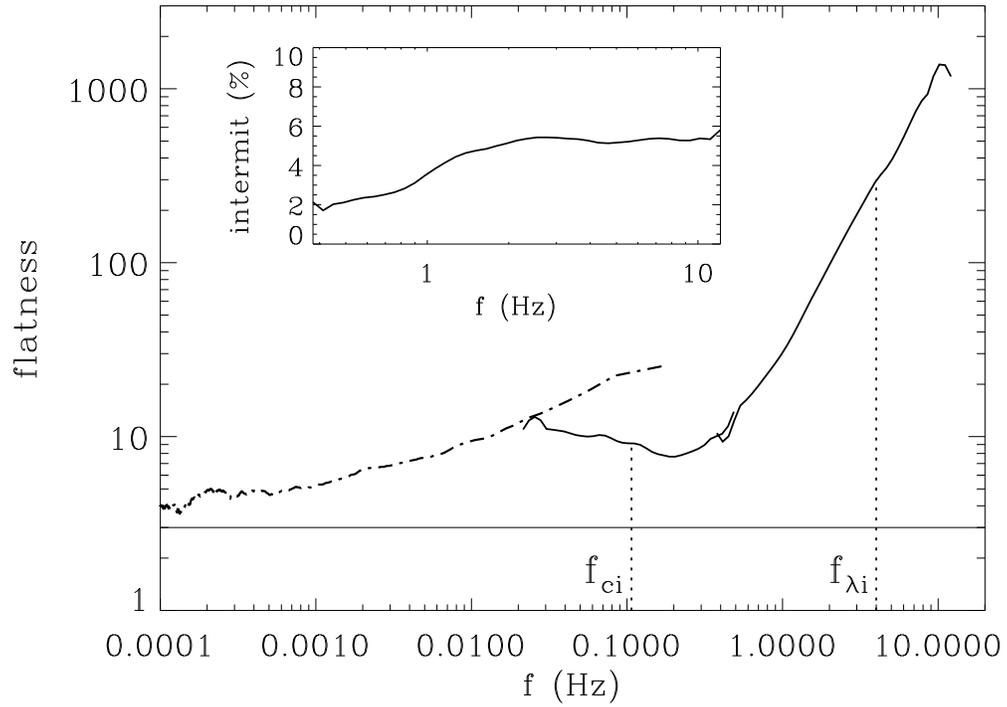}\caption{The flatness of the magnetic
field component $B_x$ calculated using Cluster data (solid line) and
Helios 2 data (dashed-dotted line); the horizontal solid line
indicates the Gaussian value $F=3$. The insert gives the percentage
of intermittent events which participate to the flatness in the high
frequency part of the spectrum.} \label{fig:flat}
\end{figure}

Intermittency may be quantified by the flatness. Fig.~\ref{fig:flat}
shows the spectrum of flatness $F(f)= \left\langle \mathcal{W}_x^4
\right\rangle/\left\langle \mathcal{W}_x^2 \right\rangle ^2$ of
$B_x$ fluctuations. We can see that in the low-frequency part $F$
increases from the large-scale Gaussian value $F=3$. This means that
the usual Alfv\'enic cascade \cite{noi} proceeds towards smaller
scales building up phase correlations among fluctuations. In the
vicinity of $f_{ci}$, Helios indicates an increase of the flatness
and Cluster/FGM shows a slight decrease. This disagreement can be
due to (i)~aliasing at the highest frequencies of Helios;
(ii)~artifacts on FGM at frequencies around the satellite spin
frequency 0.25 Hz; or to some other physical reasons which are
outside of the scope of the present paper. Here we will focus our
discussion on the higher frequencies, resolved by Cluster/STAFF-SC,
where the flatness again displays a power-law dependence on $f$.
This scaling behavior is inherent to non-linear dynamics and refers
to a new kind of "inertial region".


In non-magnetized fluid, a rapid exponential increase of the
flatness is observed in the near-dissipation range since only the
strongest fluctuations survive while the others are destroyed by
viscosity \citep{near-diss}. After such increase the flatness
saturates. In our case, the saturation is not observed and there is
a power law increase of flatness. At the same time strong coherent
structures which give an important increase of intermittency
represent only $(2-6)\%$ of the fluctuations at each scale (see the
insert of Fig. \ref{fig:flat}). This estimation has been made using
a method proposed by Farge~(1992), named \emph{local intermittency
measure} analysis. It consists in separating the intermittent events
which participate to the non-Gaussian tails of the PDF's from the
Gaussian fluctuations. At each scale $\tau$ we chose a threshold
$\zeta$ for the energy of magnetic fluctuations and we select the
wavelet coefficients which verify $|\mathcal{W}(\tau,t)|^2 < \zeta$.
With the new set of coefficients we recalculate the flatness; if it
is larger than 3 we decrease the threshold until the flatness
reaches the Gaussian value. At the end of the procedure, the
fraction of the non-Gaussian coefficients gives an estimate of the
percentage of the intermittent structures at each scale. Note, that
this analysis only gives the upper limit of the structures because
one intermittent event is described normally by more than one
wavelet coefficient.

\section{Role of compressible fluctuations}
The results described in the previous section suggest that above the
spectral break frequency $f_{b}$ there is a non-linear compressible
cascade and not a dissipation range. In this section we show that
the level of the plasma compressibility in this range can be at the
origin of a \emph{non-universality} of statistics of the magnetic
fluctuations. To show this we consider two  $7$ min time periods on
April 5, 2001, with different ion plasma $\beta_i$.

\begin{figure}
\includegraphics[angle=90,scale=.80]{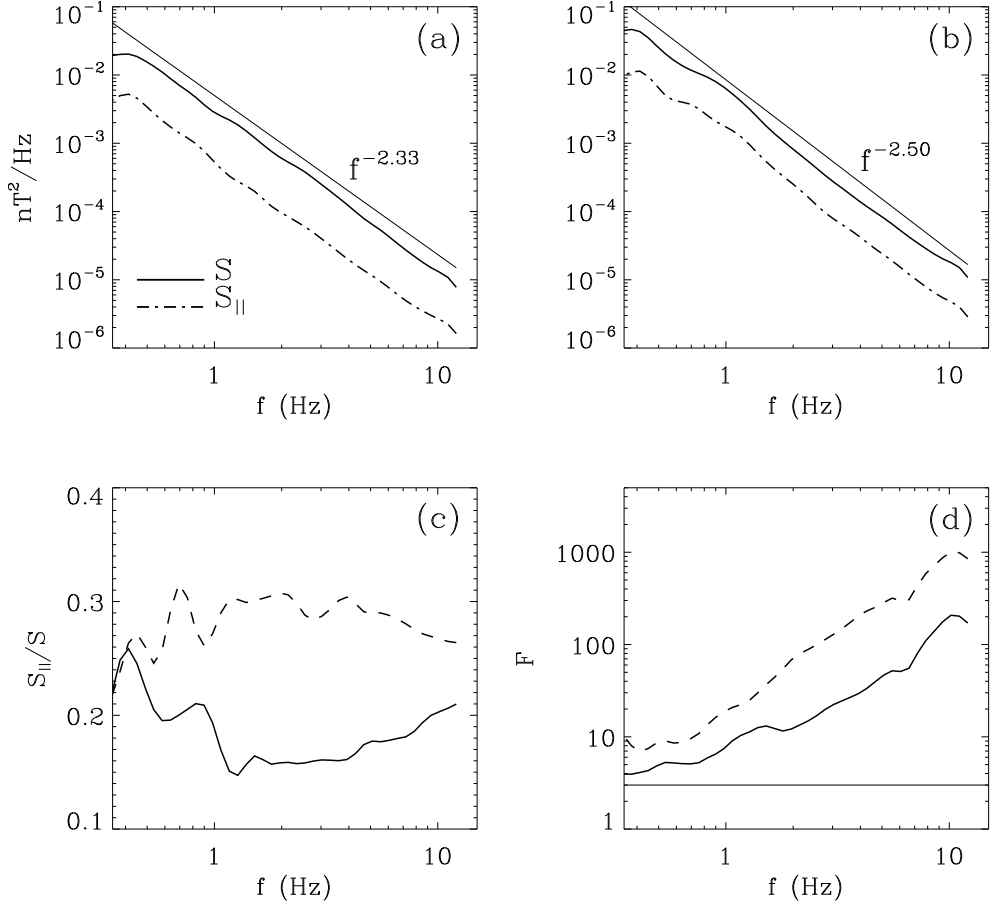} \caption{Properties of magnetic
fluctuations above the spectral break for different plasma
$\beta_i$; (a)  $\mathcal{S}$  and $\mathcal{S}_{\|}$  for a 7
minutes time period with $\beta_i\simeq 0.5$; (b) the same as (a)
but for a 7 minutes time period with $\beta_i\simeq 1.5$; (c) level
of compressible fluctuations for period (a), solid line, and (b),
dashed line; (d) flatness of the fluctuations of $B_x$ component
calculated  for the period (a), solid line, and for the period (b),
dashed line.} \label{fig:compres}
\end{figure}

Fig.~\ref{fig:compres}(a) shows the total power spectrum density of
the magnetic fluctuations within the high frequency range
$\mathcal{S}$ (solid line), and the spectrum of parallel
fluctuations of magnetic field $\mathcal{S}_{\|}$ (dashed-dotted
line) for a time period which starts at 23:03:02~UT, when
$\beta_i\simeq 0.5$. Fig.~\ref{fig:compres}(b) shows $\mathcal{S}$
and $\mathcal{S}_{\|}$ for a period starting at 22:35:02~UT, when
$\beta_i\simeq 1.5$. The straight lines indicate power law fits:
$\mathcal{S}\sim f^{-2.33}$ for the first time period and
$\mathcal{S}\sim f^{-2.50}$ for the second.
Fig.~\ref{fig:compres}(c) shows the ratio
$\mathcal{S}_{\|}/\mathcal{S}$ for the two considered time periods.
One can see that during the first time period (with $\beta_i \simeq
0.5$) the energy of compressible fluctuations is less than 20\% of
the total energy of the fluctuations (see the solid line), while for
the second, it is around 30\% (see the dashed line).
Fig.~\ref{fig:compres}(d) shows the flatness $F(f)$ calculated
within the two time periods (here again the solid line corresponds
to the period with smaller $\beta_i$ and dashed line, to the period
with higher $\beta_i$).

These observations indicate that for higher $\beta_i$, the level of
compressibility increases, leading to a steepening of the magnetic
power law and an increase of intermittency.  To confirm these
results more intervals for wider plasma parameters should be
analyzed.

\section{Nature of the small scale cascade}

How can we explain the small scale energy cascade in the solar wind
turbulence, sensitive to the compressibility of the fluctuations?
There are several interpretations. Following
\citep{Schekochihin2007}, at spatial scales between the ion and
electron characteristic scales, there is a superposition of
(undamped) Kinetic Alfv\'en Wave (KAW) and entropy cascades. KAW
waves are expected to have low frequencies $f \ll f_{ci}$ and nearly
perpendicular wave vectors with respect to the mean field,
$k_{\perp} \gg k_{\|}$.

Another possible explanation of the small scale cascade is a
superposition of non-linear fluctuations (not waves), which exchange
energy on times smaller than the usual eddy-turnover (or non-linear)
time. At spatial scales at the order of $\lambda_i$ and at
frequencies of the order of $f_{ci}$ one should take into account
the Hall effect in Faraday's equation
\begin{equation}
\frac{\partial \bB}{\partial t} = \rot \left[\bV \times \bB -
\frac{m_i}{e \rho}\bJ \times \bB + \eta \bJ \right]
\label{induzione}
\end{equation}
${\bf J}=c\nabla\times\bB/4\pi$ being the current density and $\eta$
a scalar dissipation coefficient. This equation contains three
different physical processes, with three associated characteristic
times. By introducing for each length-scale $\ell$  the mean value
of density $\rho_\ell$, magnetic intensity $B_\ell$ and velocity
$u_\ell$, we can define an eddy--turnover time $T_{NL} \sim
\ell/u_\ell$, a characteristic time associated to the Hall effect
$T_H \sim \rho_{\ell} \ell^2/  B_\ell$ and a dissipative time $T_D
\sim \ell^2/\eta$. At large scales the main effect is played by the
term $\rot (\bV \times \bB)$, which describes the low-frequency
Alfv\'enic cascade \citep{noi} realized in a time $T_{NL}$. We
conjecture that approaching the scale of the spectral break, $T_H$
appears to be of the order of $T_{NL}$ and beyond the break, the
nonlinear energy transfer is realized in a time $\sim T_H < T_{NL},
T_{D}$.

In a compressible fluid the energy balance equation must be
expressed in terms of energy densities (i.e., energy per unit
volume) and not in terms of specific energy (i.e., per unit mass),
in order to take into account density fluctuations (see
\citep{Fleck1996} for example). The mean volume rate of energy
transfer on Hall times in a compressible fluid is therefore
\begin{equation}\label{eq:e-V}
\varepsilon_V \sim \frac{\rho_{\ell} u_{\ell}^2}{T_H}.
\end{equation}

Assuming equipartition between kinetic and magnetic energies
$\rho_{\ell} u^2_{\ell} \sim B^2_{\ell}$, the energy--transfer rate
results to be proportional to
\begin{equation}\label{eq:e-V-1}
\varepsilon_V \sim \frac{B_{\ell}^2}{T_H} \sim
\frac{B_{\ell}^3}{\ell^2\rho_{\ell}}.
\end{equation}

Following von Weizs\"{a}cker (1951), the ratio of the mass density
$\rho$ at two successive levels of the hierarchy is related to the
corresponding scale size $\ell$ by the following equation
\begin{equation}\label{eq:rho-turb}
\frac{\rho_{\ell_{\nu}}}{\rho_{\ell_{\nu+1}}} = (\frac{\ell_{\nu}}
{\ell_{\nu+1}})^{-3\alpha}
\end{equation}
where $|\alpha|$ is a measure of the degree of compression at each
level $\nu$ (larger $\nu$ meaning larger length-scale), and ranges
from $\alpha=0$ for no compression up to $|\alpha|=1$ for isotropic
compression ($3|\alpha|$ is a number of dimensions in which the
compression takes place). So, using this  density scaling,
$\rho_{\ell} \sim \ell ^{-3\alpha}$, and assuming a constant
spectrum energy transfer rate we have
\begin{equation}\label{eq:b-turb}
B_{\ell} \sim \rho^{1/3} \ell^{2/3} \sim \ell^{2/3 - \alpha}.
\end{equation}
Therefore, the spectral energy function is
\begin{equation}\label{eq:E-turb}
E(k) \sim \frac{B_{\ell}^2}{k} \sim k^{-7/3 + 2\alpha}.
\end{equation}

In the incompressible plasma limit ($\alpha=0$), this phenomenology
predicts a $\sim k^{-7/3}$ spectrum. Such a spectrum has been
observed both in direct numerical simulations of an incompressible
Electron MHD turbulent system \citep{Biskamp1996} and in the EMHD
limit of the incompressible Hall MHD shell model \citep{Buchelin06}.

In the case of isotropic compression toward smaller scales
($\alpha=1$) which can take place in the interstellar medium, the
spectrum is $E(k)\sim k^{-1/3}$. If the isotropic compression is
going on toward  larger  scales ($\alpha =-1$) the spectrum will be
$E(k)\sim k^{-11/3}$. In the case, analyzed here, the power law
spectrum of magnetic fluctuations follows the $\sim k^{-2.6}$--law,
that corresponds to $\alpha \simeq -0.14$, i.e., the scaling
relation for the density is $\rho_{\ell} \sim \ell ^{0.4}$.

This simple phenomenological model allows to explain the variations
of the spectral index of the high frequency part of the solar wind
spectrum from $-4$ to $-2$ by different degree of plasma compression
with $\alpha$ between $-5/6$ and $1/6$.

However, the described model have to be improved in order to take
into account (i) the space anisotropy ($k_{\perp}\neq k_{\|}$) that
appears in a plasma with a mean field and (ii) possible different
scaling for velocity and magnetic field which can be found in the
compressible Hall MHD \citep{sergio07}. Besides, the interpretation
of the small scale energy cascade in the Hall MHD frame is supported
by the observations of a clear correlation between the spectral
break frequency $f_b$  and the Doppler-shifted wavevector of the ion
inertia length $f_{\lambda_i}$ (Leamon et al., 2000).

Note that the solar wind spectral break frequency $f_b$ is usually
observed to be lower than $f_{\lambda_i}$ \citep{Leamon98}. It can
be explained by the fact that the Hall effect starts acting at
scales larger than $\lambda_i$. In our case, for example,
$f_b/f_{\lambda_i} \simeq 0.1$ and the scale of the spectral break
is $\sim 10 \lambda_i$.

Introducing the Hall Reynolds
number as the ratio of the non-linear term over the Hall term
\begin{equation}
R_H \sim \frac{e}{c m_i} \frac{\rho_{\ell} u_{\ell} \ell}{B_{\ell}}
\end{equation}
one can see that $R_H \simeq 1$ (i.e. $T_{NL} \simeq T_{H}$) at a
scale $\ell$, that is not necessarily equal to $\lambda_i$,
$$\frac{\ell}{\lambda_{i\ell}} =4\pi \frac{V_{A\ell}}{u_{\ell}}  $$
here $V_{A\ell}$ and $\lambda_{i\ell}$ are the mean values of the
Alfv\'en speed and the ion inertial length at scale $\ell$
respectively.

\section{Conclusion}

In this paper we investigate small scales turbulent fluctuations in
the solar wind (i.e. at frequencies above the spectral break at the
vicinity of $f_{ci}$). Taking into account the Taylor hypothesis
($\ell=V/f$) the frequency domain above the spectral break covers
$\sim (40- 2000)$~km space scales that corresponds to $(0.3-
15)\lambda _i$. In this range we found evidences for strong
departure from Gaussian statistics and the presence of intermittency
while the spectrum presents a well-defined power law. Both the
presence of a power-law spectrum and the absence of global
self-similarity, seems to be quite in contrast with the role of
``dissipative range".

In usual fluid turbulence, the dissipative range  \citep{frisch}
starts with a rough exponential cutoff; in the near dissipation
range the intermittency increases as far as   the Gaussian
fluctuations dissipate faster than the coherent structures
\citep{near-diss}; then the fluctuations become self-similar, the
singularities being smoothed by dissipation. In the solar wind
turbulence we observe a completely different picture. After the
spectral break in the vicinity of $f_{ci}$ the flatness increases as
a power law indicating that non-linear interactions are at work to
build up a new inertial range.

This small scale cascade is much more compressible than the lower
frequency Alfv\'enic cascade. This sudden change in nature of the
turbulent fluctuations can happen due to a partial dissipation of
magnetic fluctuations at the spectral break: the left-hand
Alfv\'enic fluctuations with $k_{\|}\gg k_{\perp}$ are damped by the
ion-cyclotron damping \citep{Ghosh96}. Above the break, however, a new
\emph{magnetosonic cascade} takes place up to the electron
characteristic scales. This energy cascade is seems to be dominated
by the fluctuations with $k_{\perp}\gg k_{\|}$
\citep{fouad06,Mangeney2006}.

We found, as well, that the  plasma compressibility controls the
statistics of the magnetic field fluctuations. Preliminary results
show that the increase of the level of the plasma compressibility
leads to the spectrum steepening and increase of the intermittency.

To explain this small scale compressible cascade we introduce here
for the first time a simple phenomenological model based on the
compressible Hall MHD: we find that the magnetic energy spectrum
follows a $E(k) \sim k^{-7/3 + 2\alpha}$ law, depending on the
degree of plasma compressibility $\alpha$. While far from a complete
model of small scale turbulence, this simple model can explain
observed variations of the spectral index within the high frequency
part of the solar wind turbulent spectrum \citep{Leamon98,smith06}
by different degree of plasma compressibility.

\acknowledgments
We acknowledge discussions with A.~Mangeney, R. Grappin,
R.~Bruno and S.~Servidio. We thank N.~Cornilleau, E.~Lucek and
I.~Dandouras for providing Cluster data. OA acknowledges the support
of CNES and European network TOSTISP.

\clearpage

\end{document}